\documentclass{ws-procs9x6}

\begin{document}

\title{HIGH RATE PROTON IRRADIATION OF 15mm MUON DRIFTTUBES}

\author{A. ZIBELL, O. BIEBEL, R.HERTENBERGER, A. RUSCHKE, CH. SCHMITT}

\address{Fakult\"at f\"ur Physik, Ludwig-Maximilian-Universit\"at,\\
Am Coulombwall 1, 85748 Garching, Germany\\
$^*$E-mail: andre.zibell@physik.uni-muenchen.de\\
www.etp.physik.uni-muenchen.de}

\author{H. KROHA, B. BITTNER, P. SCHWEGLER, J. DUBBERT, S. OTT}

\address{Max-Planck-Institut f\"ur Physik,\\
F\"ohringer Ring 6, 80805 M\"unchen, Germany}

\begin{abstract}
Future LHC luminosity upgrades will significantly increase the amount of background hits from photons, neutrons and protons in the detectors of the ATLAS muon spectrometer. At the proposed LHC peak luminosity of 5$\cdot 10^{34}\frac{1}{cm^2s}$, background hit rates of more than 10$\frac{kHz}{cm^2}$ are expected in the innermost forward region, leading to a loss of performance of the current tracking chambers.

Based on the ATLAS Monitored Drift Tube chambers, a new high rate capable drift tube detecor using tubes with a reduced diameter of 15mm was developed.

To test the response to highly ionizing particles, a prototype chamber of 46 15mm drift tubes was irradiated with a 20 MeV proton beam at the tandem accelerator at the Maier-Leibnitz Laboratory, Munich.
Three tubes in a planar layer were irradiated while all other tubes were used for reconstruction of cosmic muon
tracks through irradiated and non‐irradiated parts of the chamber. To determine the rate capability
of the 15mm drift‐tubes we investigated the effect of the proton hit rate on pulse height, efficiency and spatial
resolution of the cosmic muon signals.
\end{abstract}


\bodymatter

\section{Introduction}\label{aba:sec1}
At the propsed LHC peak luminosity of 5$\cdot 10^{34}\frac{1}{cm^2s}$, the innermost foreward region of the ATLAS muon spectrometer will be confronted with a background hit density of about 10 $\frac{kHz}{cm^2}$, as the rates scale with the luminosity\cite{radtask}, leading to a loss of performance of the current tracking chambers caused by high occupancy and space charge effects.

An upgrade candidate are MDT\footnote{\label{foot:MDT}Monitored Drift Tube}-chambers with a reduced tube diameter of 15 mm\cite{bittner}. To study the high-rate capability, a prototype chamber with 46 tubes was irradiated by 20 MeV protons at the MLL Tandem-accelerator in Garching.

\section{Experimental setup}
\begin{figure}
	\centering
	\psfig{file=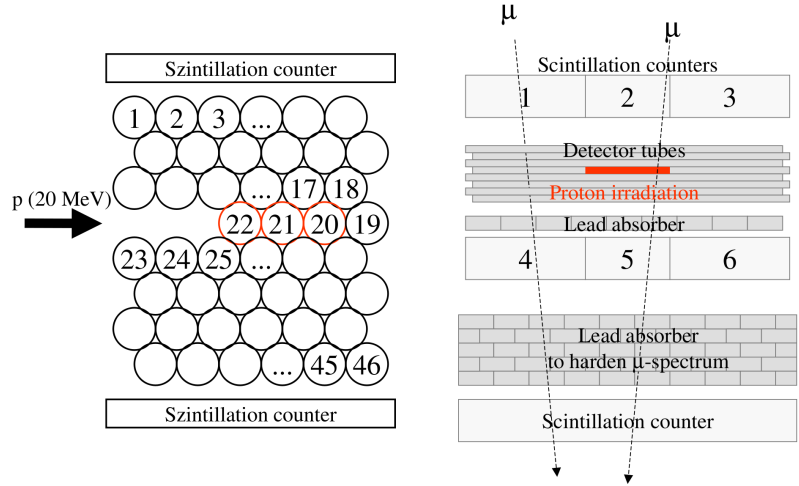,width=4.3in}
\caption{Schematic side-view of the detector and view from beam direction. The lead absorber and the additional trigger scintillator are used to harden the muon spectrum.}
\label{fig:exp_setup}
\end{figure}

Figure~\ref{fig:exp_setup} shows the experimental setup.
A fast coincidence of the scintillation counters serves as trigger on cosmic muons.
The proton beam was defocussed to a beam-spot of 3$\cdot$0.5 $cm^2$ and wobbled with 800 Hz over a horizontal distance of 7 cm.
Only the tubes 20 - 22 were irradiated by protons, that are - consistent with SRIM\footnote{\label{foot:SRIM}The Stopping and Range of Ions in Matter, http://www.srim.org} simulations - stopped by the second or third tube wall, depending on small angle scattering in the first two tubes.

The triggering scintillators are segmented, thus one can distinguish between muons crossing the irradiated and non-irradiated sections along the 1 m tubes, see fig.~ \ref{fig:exp_setup}.
Three different beam intensities were used during the measurements, 200 kHz, 1100 kHz and 1300 kHz, corresponding to an effective hit-density of 19, 105 and 124 $\frac{kHz}{cm^2}$. Reference runs with no beam were taken.

\section{Signalheight}
\begin{figure}
	\centering
	\psfig{file=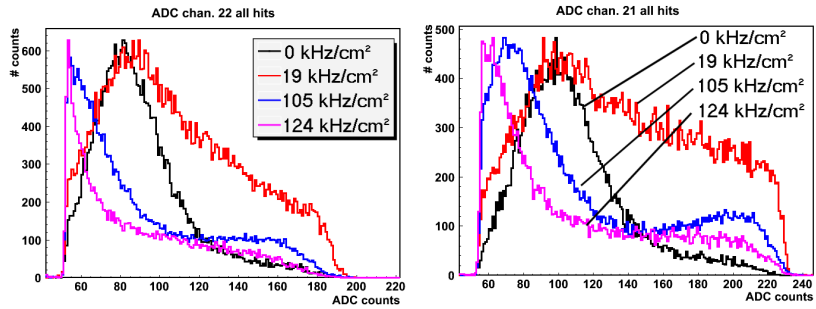,width=4.3in}
\caption{Pulse height spectrum of the first (left) and second (right) irradiated tube at different irradiation levels.}
\label{fig:adc_all}
\end{figure}

The MDT electronics provides information on drifttime (time between trigger and anode wire signal to reconstruct the distance of the track from the wire) and signalheight of the analog pulse from the wire.
This signalheight is measured in ADC-counts, and the signal height spectra for the first two tubes in the irradiated detector layer are shown in fig.~\ref{fig:adc_all} for the different proton irradiation levels.

In the case of no proton irradiation, all four tubes show a similar spectrum with Landau like energy distribution. At 19 $\frac{kHz}{cm^2}$ kHz proton irradiation, the tubes 22 and 21, being the first ones hit by the beam, show an overlay of the muon signals with artificial proton coincidences. The energy spectrum of these proton hits is very broad at high values, even above the dynamic range of the ADC, where the spectrum is cut off.
At the highest irradiation levels, the ADC spectra of tubes 22 and 21 are shifted to the left, where they are cut off by the ADC threshold for small signals. This is due to the reduction of the gas amplification close to the anode wire, caused by the positive ions, drifting slowly to the tube walls.
The electric field of these space charges shields the area around the wire and reduces the effective electric field for gas amplification.
Protons reaching the gas volume of the third tube are stopped there, depositing several MeV of energy. As the impact on the ADC spectrum is different, this tube is excluded from further consideration.
Tube 19 shows identical ADC spectra for all irradiation levels, as all protons are stopped before entering this tube.

\begin{figure}
	\centering
	\psfig{file=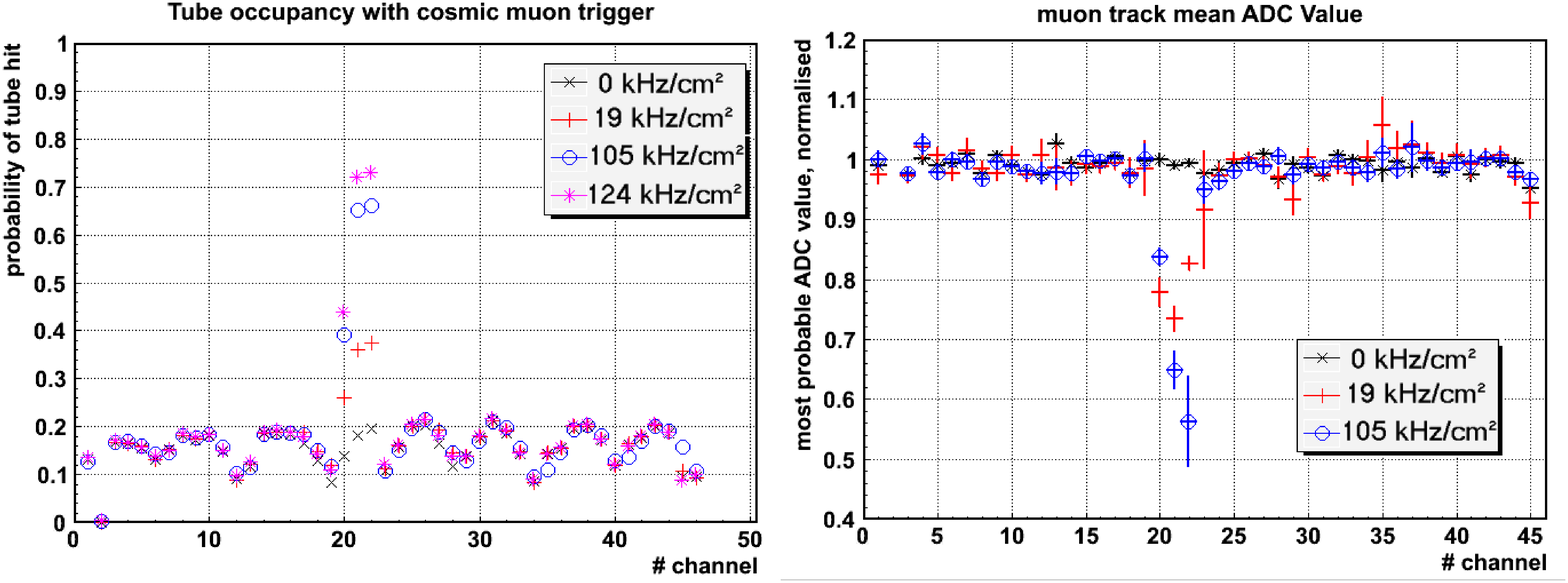,width=4.3in}
\caption{Tube occupancy including muon and proton hits (left) and mean ADC values for muon tracks through the tubes (right) at different irradiation levels.}
\label{fig:occu}
\end{figure}

Figure~\ref{fig:occu} (left) shows the probability of a detected signal in each tube for a muon trigger. In the case of no irradiation, the acceptance of the detector can be observed, with its periodical 8-layer structure with lower acceptance at the chamber edges. Due to random coincidences of a proton hit with a cosmic muon, at 200 kHz there is a 1:1 weighting of muon and proton hits in the irradiated layer, that reaches a ratio of up to 2.5:1 at 1300 kHz irradiation.

If one considers the most probable ADC value for hits, that could be matched to tracks of cosmic muons through the respective tube, this can be taken as an indicator for the amount of gas amplification. Figure~\ref{fig:occu} (right) shows these values for all tubes and irradiation levels, normalised to their values at non-irradiated sections of the tubes, except for 1300 kHz, where the lack of statistics did not allow this analysis.

\section{Efficiency and Tracking resolution}

The spatial resolution of a drifttube can be obtained by comparing its radius prediction with the one derived from the track fit through the non-irradiated tube layers (all beside layer 4).
Comparing these two different radii introduces the 3-sigma-efficiency, giving the fraction of tracks where this difference lies within 3 times the spatial resolution of the tube.

\begin{figure}
	\centering
	\psfig{file=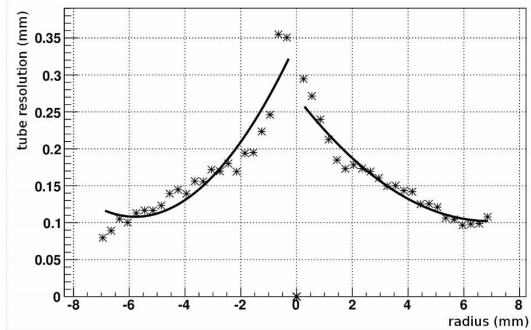,width=3in}
\caption{Spatial resolution for non irradiated tubes.}
\label{fig:resolution_efficiency}
\end{figure}

Radius dependent resolution is shown in fig.~\ref{fig:resolution_efficiency}. The resolution is predicted to deteriorate only by a few ten $\mu$m due to space charge effects under irradiation\cite{bittner}. Efficiency drops at 19 $\frac{kHz}{cm^2}$ due to space charge effects in the irradiated section and due to occupancy effects, dependent on the total hit rate per tube.

The exact analysis depends strongly on the definition of the 3-sigma-efficiency, this is still ongoing work.

\section{Conclusion}
With the Garching Tandem accelerator, the threefold of the predicted values for background irradiation with highly ionizing particles at high-luminosity-LHC could be simulated.
Due to dead-time and space charge effects the tracking efficiency and gas amplification is reduced.

These effects are small compared to detectors of the currently used drifttubes with 30mm diameter, where the occupancy is about seven times higher, and the degradation of efficiency and resolution is much worse.
The number of tube layers per 15mm MDT chamber will be doubled compared to a 30mm tube chamber. Thus the overall segment tracking efficiency of the system at 19 $\frac{kHz}{cm^2}$, about twice the expected hit-rate at high luminosity LHC, meets the requirements for the forward region of the ATLAS muon spectrometer.

\end{document}